\documentstyle[emulapj]{article}

\begin{document}

\font\tenbg=cmmib10 at 10pt
\def \rvecxi{{\hbox{\tenbg\char'030}}}
\def \rvecphi{{\hbox{\tenbg\char'036}}}
\def \rvecdelta {{\hbox {\tenbg\char'016}}}
\def \rvecepsilon {{\hbox {\tenbg\char'017}}}
\def \rvecmu{{\hbox{\tenbg\char'026}}}
\def \rvecOmega {{\hbox {\tenbg\char'012}}}

\title{MHD Simulations of Disk-Magnetized
Star Interactions
   in Quiescent Regime: Funnel Flows
and Angular Momentum Transport}

\author{M.M. Romanova \altaffilmark{1},
G.V. Ustyugova \altaffilmark{2},
A.V. Koldoba \altaffilmark{3}, and
R.V.E. Lovelace \altaffilmark{4}}

\altaffiltext{1}{Department of Astronomy,
Cornell University, Ithaca, NY 14853-6801;
Romanova@astrosun.tn.cornell.edu}

\altaffiltext{2}{Keldysh Institute of Applied Mathematics,
      Russian Academy of Sciences, Moscow, Russia;
Ustyugg@spp.Keldysh.ru}

\altaffiltext{3}{Institute of Mathematical Modelling,
      Russian Academy of Sciences, Moscow, Russia;
Koldoba@spp.Keldysh.ru}

\altaffiltext{4}{Department of Astronomy,
Cornell University, Ithaca, NY 14853-6801;
RVL1@cornell.edu }

\begin{abstract}

Magnetohydrodynamic (MHD) simulations have been used
to study disk accretion to a rotating magnetized star
with an aligned dipole moment.
      Quiescent initial conditions were
developed in order to avoid the fast initial
evolution seen in earlier studies.
      A set of simulations was performed for different stellar
magnetic moments and rotation rates.
Simulations have shown that the disk structure
is significantly changed inside
a radius  $r_{br}$ where magnetic braking is
significant.
     In this region the disk is strongly inhomogeneous.
     Radial accretion of matter slows as it approaches the
area of strong magnetic field and a dense ring and funnel flow
form at the magnetospheric radius $r_m$ where the magnetic
pressure is equal to the total, kinetic plus thermal,
pressure of the matter.

     Funnel flows (FF), where the disk matter
moves away from the disk plane and flows along
the stellar magnetic field, are found to be
stable features during many rotations of the disk.
     The dominant force
driving matter into the FF is the
pressure gradient force,
while gravitational force accelerates it
as it approaches the star.
     The magnetic force is much smaller
   than the other forces.
     The funnel flow is found
to be   strongly sub-Alfv\'enic everywhere.
      The FF is subsonic
close to the disk, but it becomes
supersonic well above
the disk.
Matter reaches the star with a velocity
close to that of free-fall.

      Angular momentum is
transported to the star
dominantly  by the magnetic field.
      In the disk the transport of angular
momentum is  mainly by the matter,
but closer to the star the
matter transfers its angular momentum to the
magnetic field
and the magnetic field is dominant in
transporting  angular
momentum to the surface of the star.
      For  slowly rotating stars we observed that
magnetic braking leads to the deceleration
of the inner regions of the disk
and the star spins up.
      For a rapidly rotating
star,  the inner regions of the disk
rotate with  a super-Keplerian velocity, and the
star spins-down.
    The average  torque is found to be zero when the
corotation radius $r_{cor}\approx 1.5 r_m$.

  The evolution of the  magnetic field  in the corona
of the disk depends on the ratio of magnetic to matter
energies in the corona and in the disk.
     Most of the simulations
were performed in the regime of a relatively
dense corona where the matter energy
density was larger than the magnetic energy density.
     In this case the coronal magnetic field
gradually opens but the velocity and density of
outflowing matter are  small.
     In a test case where a
significant part of the corona was in
the field dominated regime,
more dramatic opening of the magnetic
field was observed
with the formation of magneto-centrifugally
driven outflows.

    Numerical applications
of our simulation results are made to
   T Tauri stars. We conclude that our quasi-stationary
simulations  correspond to the classical T Tauri stage
of evolution.
    Our results
are also relevant to cataclysmic
variables and magnetized neutron
stars in X-ray binaries.

\end{abstract}


\section{Introduction}

        Disk accretion to a rotating magnetized star
is important in a number of astrophysical objects,
including T Tauri stars
(Edwards  {\it et al.} 1994), cataclysmic variables
(e.g., Warner 1995), and X-ray pulsars
(e.g., Bildsten  {\it et al.} 1997).

\begin{figure*}[t]
\centering
\caption{Initial conditions of Types I (left panel) and
II (right panel). The gray-scale and numbers show
density distribution.}
\label{Figure 1}
\end{figure*}

The accretion of matter to a rotating
star with a dipole magnetic field is
a complicated problem still only partially solved.
     The important questions which  need to be answered include:
(1) What  is the structure of the disk near the magnetized star?
(2)  Where is the inner radius of the disk?
(3) What is the nature of the funnel flows (FF)?
     For example, which force is
dominant in lifting matter to the funnel flow?
(4) How is the accretion rate influenced by the star's magnetic
moment $\mu$ and angular velocity $\Omega_*$?
(5) What is the mechanism of angular momentum transport between the
star and the disk?
  What determines whether star spins-up or spins-down?
(6) What are the necessary conditions for
magneto-centrifugally  driven outflows from the disk and/or
the star?

Many   of these questions have been investigated analytically,
but the  conclusions reached by different authors often
differ  because the simplifying assumptions are
different.
    For example, regarding question (2),
some authors conclude that the disk should be disrupted
  in the region where magnetic and matter stresses
are comparable (e.g., Pringle and Rees 1972 - hereafter PR72;
Davidson \& Ostriker 1973;
Lamb, Pethick \& Pines 1973;
Ghosh, Lamb, \& Pethick 1977; Scharlemann 1978;
Ghosh \& Lamb 1979 a,b - hereafter GL79a,b;
Camenzind 1990; K\"onigl 1991; Shu {it et al.} 1994 --
hereafter S94).
    Others, argue that the inner radius of the disk
should be farther away, at the corotation radius,
because the inner regions of the disk are disrupted
by magnetic braking
  (Ostriker \& Shu 1995 -- hereafter OS95;
Branderburg \& Campbell 1998;
Elstner \& R\"udiger 2000).

\begin{figure*}[t]
\centering
\caption{Evolution of a disk with no magnetic
field for initial conditions of Type I (panels a, b, c, d)
and Type II (panels e, f, g, h). Panels a \& e show initial
density distribution.
    Other panels show
density distribution for different viscosities,
  $\alpha=0, ~0.01,$ and $ 0.02$ at $T=50$.
       The initial density for initial conditions of Type II is about 1.5
times smaller than for Type I.
Consequently the density scale is slightly
different in these cases.
   The  legend for $\rho$ is chosen to show
the density distribution
in the main part  of the disk.
    It does not reflect the low density
in the corona where $\rho_{min}=0.003$
and the highest densities of matter in the disk
  near the star where $\rho_{max}=18$ at the panel $c$,
$\rho_{max}=12$ at the panel $d$,
$\rho_{max}=2.1$ at the panel $g$,
$\rho_{max}=4.2$ at the panel $h$.}
\label{Figure 2}
\end{figure*}

   Question (3) is
investigated in only a few papers which
use the Bernoulli integral
(Lamb, Pethick \& Pines 1977;
  Paatz \& Camenzind 1996 -- hereafter PC96;
Li \& Wilson 1999; and
Koldoba {\it et al.} 2002).
    The authors agree that the flow
should become supersonic (and slow magnetosonic)
just above the disk.
   On the other hand opinions differ
regarding the driving force which pushes
matter up into the FF.
Li and Wilson (1999) (see also Li et al. 1996)
propose that the twisting of the
  magnetic field near the base of the FF
should be very large,
$\gamma_\phi=|B_\phi|/B_p >> 1$ and the magnetic force should
be the main one lifting matter to the FF.  Here,  $B_\phi$
is the toroidal component of the magnetic field and $B_p$
is the poloidal component.
      Other groups
  (e.g., Lamb et al. 1977; PC96;
Koldoba et al. 2002) argue
that the FF should be  super-Alv\'enic
so that the twisting of the field is small
so that the magnetic force is also small.
In numerical simulations by
MS97, H97 and GBW99 the magnetospheric accretion was reported,
 but no clear evidence of funnel flows
was presented and no analysis of FFs was performed.
  A significant part of this paper is devoted to the FFs.
       Another important issue which has been
discussed over the past $30$ years is question (5)
concerning the transport of angular momentum
between the disk and the star.
    What determines the sign of the torque
on the star?
     In early papers it was
supposed that a star can  only be spun-up because
matter in a Keplerian disk brings positive angular momentum to
the star (e.g., PR72).
     Later, it was recognized that a star can be
spun-down due to the part
of the star's magnetic flux which
passes through the disk outside of
the corotation radius
(GL79b, Wang 1995).
     Recently, the idea of
``torqueless'' accretion was proposed
where mass but not
  angular momentum
is transported to a star (e.g. S94;
OS95; Li, Wickramasinghe \& R\"udiger 1996;
Li \& Wickramasinghe 1997). Wang (1997) presented
arguments against this idea, but this still remains
an open question.

     Question  (6) regarding
  magneto-centrifugally
driven outflows from the disk
has been discussed by a number of authors
(e.g., Camenzind 1990; K\"onigl
1991;  S94;
OS95;  Lovelace, Romanova \& Bisnovatyi-Kogan 1995 -
hereafter LRBK95;
  Fendt, Camenzind \& Appl 1995;
PC96; Goodson \& Winglee 1999;
Bardou \& Heyvaerts 1999;
Agapitou \& Papaloizou 2000).
     For example,  S94 proposed
that poloidal magnetic flux  accumulates
near the corotation radius
and magnetic winds should blow from this point (X-point).
LRBK95 proposed that wind may form from the entire region
of the disk outside the corotation radius where
the magnetic field threading the disk is open.

\begin{figure*}[t]
\centering
\caption{Evolution of the disk and the
poloidal magnetic field in the
largest region studied, $R_{max}=55$.
    The density (gray scale background)
varies from a minimum value $\rho=0.003$ in the corona
to a maximum value $\rho=2.6$ in the disk.
The field lines are labeled by their magnetic flux $\Psi$ values
which  change from $0.0009$ to $4.9$.}
\label{Figure 3}
\end{figure*}

    Analytical investigations of disk accretion
to a magnetized star are of course limited by
the different assumptions made.
    For this reason robust 2D and 3D
simulations are essential to further the understanding
of the different phenomena.
    By robust we
mean that the result should not depend on initial
conditions, boundary conditions,  on grid resolution, and
other  artificial factors.
      Several 2D MHD simulation studies have been made
with different initial conditions
  aimed at disk/star outflows.
       In an early work Hayashi, Shibata, and Matsumoto (1996)
(hereafter HSM96) investigated the interaction of a non-rotating
star with a Keplerian accretion disk
and observed the opening of magnetic field lines
which initially thread both the star and the disk.
    They found single event
outflows and the corresponding inward collapse
of the disk on a dynamical time-scale
  (less than one period of rotation
of the inner radius of the disk)
with the radial  velocity of
the disk close to free-fall.
    This fast
evolution was the result of the magnetic braking
of the disk by the  magnetic field linking the
disk and non-rotating star through a conducting corona.
     This explosive behavior may correspond to some
episodic accretion events of actual systems.
     However, it is important to investigate
the possible quiescent behavior of the disk-star
systems.

Miller and Stone (1997; hereafter MS97)
investigated disk-star interaction
for different geometries and stellar
magnetic fields
using the resistive ZEUS code.
    MS97 rotated the corona - which
occupies all the space between the
star and the disk -
with the rotation rate of the star.
     This
decreased the initial magnetic
braking (compared to HSM96), so that they were
able to perform simulations for
several periods of rotation of the inner radius of the disk.
    In cases with a relatively weak magnetic field, they
got results similar to those of HSM96.
     They also found the disk
collapsing to the star with velocity
$\sim 0.5 v_{ff}$, the opening
of magnetic field lines,
  and outflows of matter from the disk.
     However, in the case of a strong
magnetic field, particularly in the
case which included   a uniform
homogeneous vertical magnetic field threading
the disk,
they observed diminished outflows.
    Instead the  matter flowed
around the magnetosphere to the star.
Similar results were obtained by
Hirose et al. (1997; hereafter H97).

Goodson, Winglee and B\"ohm (1997)
(hereafter GWB97) and Goodson, B\"ohm
and Winglee (1999) (hereafter
GBW99) did much longer simulations
in very large simulation regions.
      They observed quasi-periodic
matter outbursts   associated with the
quasi-periodic opening of magnetic field lines
and matter accretion to the star.
The density in the corona was chosen to
decrease in a special way, $1/R^{4}$,
so that the Alfv\'en speed $\propto B/\sqrt\rho
\propto 1/R$ decreases gradually.
    This is  favorable for the
opening of magnetic field lines, and for the
generation and propagation of outflows.
    GWB97 and GBW99 do not investigate cases
where the density falls off more slowly with distance.


\begin{figure*}[t]
\centering
\caption{Evolution of the disk and the magnetic field in a
medium size region $R_{max}=7$.
      The configurations
at $T=0, 10, 20, 30, 40, \& 50 $ are shown.
   The density (gray scale background)
changes from maximum value $\rho=2.6$ in the disk to
the minimum value $\rho=0.003$ in the corona.
    Contour levels of $\Psi$, which label
poloidal field lines, change from $4.9$ to $0.006$.
The field lines of the strongest magnetic field near the star
are not included in order to make
the inner structure of the disk visible.}
\label{Figure 4}
\end{figure*}

Fendt and Elstner (1999, 2000 -- hereafter FE99, FE00)
investigated disk-star interaction
for thousands of rotations of the inner radius of the disk,
and observed the opening of magnetic field lines and outflows.
    However, they treated the disk as a boundary condition
so that they could not take into account the back reaction
of the disk on the stellar magnetic field.
     Furthermore, the actual outflow
of matter from the disk to the corona may
be different from that assumed.

     The above -- mentioned simulation studies show that
outflows appear either in very non-stationary
situations (HSM96, MS97) or for a special
distribution of  coronal density and very fast rotation
of the star (GWB97, GBW99).
None of the  papers give answer to questions
(1)-(5).
      Also, it is not clear whether or not outflows
exist for more quiescent
initial conditions, and for cases where the
coronal density falls off slowly with distance.

    In this paper we investigate disk accretion to a
rotating magnetized star and the associated funnel flows.
    We start from
initial conditions which give us the possibility to significantly reduce
the initial magnetic braking between the disk and  the corona.
    This allows us
to investigate the disk-star interaction and funnel flows
for long times and to consider questions (1)-(5)
in detail.

    In \S 2 we describe the numerical model, including
initial and boundary conditions.
    We also discuss the evolution of the disk
without a magnetic field.
     In \S 3 we describe in detail the disk-star
interaction for the case of  slowly rotating stars
and in  \S 4 cases of fast rotating stars.
In \S 5 we consider the dependence of disk-star interaction
on the magnetic moment $\mu$.
    In \S 6 we analyze the physics of FFs.
In \S 7 we consider the possibility of outflows.
In \S 8 we apply our simulation results to T Tauri stars.
    In \S 9 give the conclusions from this work.

\begin{figure*}[t]
\centering
\caption{Evolution of the density
in the disk (color background)
and poloidal magnetic field lines  ($\Psi=$const lines)
(white lines)  in case of
Type I initial conditions in the region $r< 6$.
    The black solid line corresponds to $\tilde\beta = 1$.
The density changes from $0.003$ in the corona to
$2.4$ in the disk.
   Contours of $\Psi$ are exponentially
spaced between $0.2$ and $1.2$.}
\label{Figure 5}
\end{figure*}

\begin{figure*}[t]
\centering
\caption{Same as on Figure 4 but for initial conditions
of Type II.
At $T=0$ the inner radius of
the disk is at $r=r_{cor}=3$.
The $\Psi=$const lines are exponentially spaced between
    $0.1$ and $0.7$.}
\label{Figure 6}
\end{figure*}

\section{Numerical Simulations}

To investigate the disk-star interaction,
we numerically solve the  MHD
equations,
$$
{\frac{\partial \rho}{\partial  t}}  +
{ {\bf \nabla}\cdot (\rho {\bf v})} =  0~~{\rm (1)},~~
{\frac{ \partial (\rho {\bf v})}{\partial  t}}  +
{{\bf \nabla}\cdot {\cal T}}  =  \rho {\bf g}~~~~
\eqno(2)
$$
$$
{\frac{ \partial {\bf B}}{\partial  t}} -
{{\bf\nabla}\times ({\bf v}\times {\bf B})}
= 0~~~{\rm (3)},~~
{\frac{ \partial( \rho S)}{\partial  t}} +
{{\bf \nabla}\cdot (\rho  {\bf v} S)}  =  0~.
\eqno(4)
$$
Here, $S$ is entropy,
     ${\cal T}_{jk}
=  p \delta_{jk}+\rho v_j v_k
  + ( {\bf B}^2
\delta_{jk}/2 -$ $B_j B_k)/(4\pi)$
is the stress tensor,
${\bf g} = - {\bf \nabla} \Phi$
is the gravitational acceleration, and
$\Phi=-GM/R$ is the gravitational
potential of the central object.
      We take the equation of state to be
$S = p/\rho^\gamma$,
where $\gamma = 5/3$ for
most of this work.
We solve equations
(1)-(4) in a spherical coordinate
system $(R,\theta,\phi)$.

      In order to model the slow accretion of
matter in the disk we modified equations
(2) and (4) by including a small
viscous stress $\sigma_{jk}$
in  ${\cal T}_{jk}$.
       Because the flow is dominantly in
the azimuthal direction the main
viscous stress acts in
this direction.
       Thus the important viscous
stress components are the
$(R\phi)$ and
$(\theta\phi)$ terms,
$\sigma_{R\phi}=-\nu
\rho R \sin \theta {\partial\omega}/{\partial R}$,
$\sigma_{\theta\phi}=-\nu \rho  \sin \theta
{\partial\omega}/{\partial \theta}$,
where $\omega=v_\phi/{R\sin
\theta}$ is angular velocity of the matter, and
$\nu$ is the kinematic viscosity.
We adopt the
$\alpha-$ model for the viscosity
(Shakura \& Sunyaev 1973),
where $\nu = \alpha c_s^2/\Omega_K$, with
$c_s=(p/\rho)^{1/2}$  the
isothermal sound speed velocity and
  $\alpha \ll 1$ is a dimensionless parameter, which
was chosen to be $\alpha=0.01$ or $\alpha=0.02$
   in most of simulations
which gave a reasonable speed of matter flow to the
central regions.
      We checked that for the  grid sizes used
the numerical viscosity
is much lower than the considered
$\alpha$ viscosity (see \S 2.5).

Equations (1) - (4)
were solved with a Godunov
type numerical code developed and
tested by Koldoba {\it et al.} (1992),
Koldoba and Ustyugova (1994) and
Ustyugova {\it et al.} (1995).
      This type of  code
has also been  described
by Ryu, Jones, and Frank (1995).
    A modified version of
the code incorporating viscosity
was tested in a number of hydrodynamic simulations
for different values of $\alpha$,
from $\alpha=0.01$ to $\alpha=0.1$.
    For $\alpha \sim 0.01- 0.05$ we observed a steady
flow of matter to the star with small velocities
(see \S 2.5)  corresponding approximately to the
Shakura-Sunyaev accretion model.

  Our Godunov type code does not explicitly include
magnetic diffusivity
 (unlike  the codes of  MS97,
H97, \& GBW99). 
    However, 
 matter  diffuses across the magnetic field
 owing to the small numerical diffusivity $\nu_m$
 which is of the order
of the numerical viscosity $\nu_m\sim \Delta R c_s$,
where $\Delta R$ is the grid spacing.
   The estimated magnetic Reynolds number associated with the
radial flow is $Re_m \sim
\delta R v_r /\nu_m$,  where $\delta R$  is the
characteristic scale.
   In the region $1 < R < 5$,
$Re_m\sim 3$ owing to the small radial velocity $v_r$.
   In actual accretion disks, both angular
momentum transport and diffusion of the magnetic field,
are probably due to turbulence 
(e.g., Bisnovatyi-Kogan \& Ruzmaikin 1976;
Balbus \& Hawley 1991) and have similar order
of magnitude transport coefficients.
    In our simulations we observed that
both the viscosity and the
magnetic diffusivity are small and are of the same
order of magnitide. 
   Note, that in the $\phi$ direction, the
velocity of the flow is high, and the Reynolds number
is also high,
$(Re_m)_\phi \sim 100$, so that magnetic field lines
rotate with the disk in $\phi$ direction, and at the same
time matter slowly diffuses in $r-$direction.
This is important for understanding of evolution of
magnetic field lines observed in the following sections.

\subsection{Reference Values and Dimensionless Units}

        A reference value of distance is
denoted $R_0$.
       The reference value for velocity is
taken to be $v_0 = (GM/R_0)^{1/2}$
which is the Keplerian velocity at
$R_0$.
      The reference angular rotation
rate is $\omega_0 = v_0/R_0$, and the
corresponding time-scale is $t_0=R_0/v_0$.
         For discussing our results
we also use the rotation period  at $r=R_0$,
$P_0=2\pi t_0 $.
       For a given magnetic field strength
at $R_0$ we can define a reference
density as $\rho_0 = B_0^2/v_0^2$
and reference pressure $p_0=\rho_0 v_0^2$.
          A reference magnetic moment of
the star is then $\mu_0 =B_0 R_0^3$.
Thus, the calculated variables are:
$R'=R/R_0$, $t'=t/t_0$, $\rho'=\rho/\rho_0$, $v'=v/v_0$,
$B'=B/B_0$, $p'=p/p_0$.
     We also use the dimensionless time,
  $T=t/P_0$.
    Later, we omit the primes but note
that any dimensionless variable
can readily  be converted to its physical value.

     Here, we give parameters for a
typical T Tauri star.
    We take the mass and radius
of the star to be
$M_*=0.8 M_\odot$ and
$R_*=2 R_\odot$.
    We use as the length scale $R_0$  the initial
inner radius of the disk, so that the dimensionless
inner radius of the disk is $R_d=1$
in most of simulations. We place the inner boundary (a star)
at $R_{min}=0.35$.
         Then, our reference length is
$R_0\approx  R_*/0.35 \approx
  4.0\times 10^{11}~{\rm cm}$, while the
simulation region corresponds to
$R_{max}=55\times R_0\approx 1.47~{\rm a. u.}$.

       The reference  velocity  is
$v_0 \approx 1.63\times 10^7~{\rm cm/s}$,
and the corresponding time-scale is
$t_0\approx R_0/v_0 \approx 2.45\times 10^4~{\rm s}$.
     The period of rotation of the inner radius
of the disk is
$P_0 \approx 1.78~ {\rm days}$.

     Consider a magnetic  field strength
$B_*=10^3~{\rm G}$ at the star's surface.
    Then at
$R=R_0$, the reference magnetic field
is  $B_0=B_*(R_*/R_0)^3 \approx 42.9
~{\rm G}$ and the reference magnetic moment is
$\mu_0=B_0 R_0^3 \approx
2.7\times 10^{36}~{\rm G cm^3}$.
        The reference density is
$\rho_0=6.9\times 10^{-12}~{\rm g/cm^3}$,
or $n = 6.93\times 10^{12}~{\rm cm^{-3}}$ which is typical
for T Tauri star disks.
The reference mass accretion rate is
$\dot M_0 = \rho_0 v_0 R_0^2 \approx
1.8\times 10^{18}~{{\rm g/s}} \approx
2.8\times 10^{-7}~
{\rm M_\odot}/{\rm year}$.
The reference angular momentum flux
is: $\dot L_0= \rho_0 v_0^2 R_0^2$.

\begin{figure*}[b]
\centering
\caption{Radial distribution of angular velocity
along the equatorial region of the
disk for $r \leq 2$
    at different times $T$.
The solid line shows the Keplerian velocity.}
\label{Figure 7}
\end{figure*}

\subsection{The Grid}

  We place the inner boundary of the simulation
region - the ``star'' - at $R_{min}=0.35$. The size of
external boundary depends on the grid (see below).
     The spherical grid was inhomogeneous
in the $R-$direction.
      The inhomogeneity
was such that cells at any distance $R$
were  approximately  square with
$\Delta R \sim R \Delta \theta$ for
fixed $\Delta \theta$.
       This grid gives a high space
resolution close to the star
which is important in this problem.
At the same time simulations in very large
regions may be performed. The inhomogeneous
grid is a smooth analog of the nested grids
used by GBW97 and GBW99.

     The angular grid was taken to have
$N_\theta=51$ in most cases,
and $N_\theta=71$ in  cases with very strong magnetic field
(see \S 7).
     The number of points in the $R-$direction determined
the size of the simulation region.
     The main  runs were done with a
  large radial grid  $N_R=150$ which corresponds to
$R_{max}=55 \approx 157 R_{min}$.
     Simulations
with a smaller grid  $N_R=100$    corresponding to
$R_{max} = 10\approx 35 R_{min}$ were done in a number of cases.
     The large size of the region was chosen in order
to minimize the influence of
external boundaries on
  processes in the vicinity of the star.
      We checked, however, that results of
  simulations in the larger region almost coincide with those
in the smaller region.
      For the typical grid with
$N_\theta=51$, the smallest cell near the star corresponds
to  $\Delta R = 0.01\approx 0.03 R_{min}$ which is
a resolution about $3.2$ times higher than one in
  GBW99.
     The largest grid is at the external boundary,
for $N_R=150$ and is $\Delta R = 5.4 R_{min}$.
    These numbers
show that a spherical stretched grid
is advantageous for
simulations of accretion flow to a star with a
dipole field.

\subsection{Boundary Conditions}

       At the inner boundary  $R=R_{min}$,
we apply  ``free" boundary
conditions for the density, pressure, and entropy,
${\partial\rho}/{\partial R}=0$,
${\partial p}/{\partial R}=0$, and
${\partial S}/{\partial R}=0$.
      The inner boundary is treated as a
rotating perfect conductor
${\bf \Omega}=\Omega \hat{\bf z}$, which
is analogous to the case of
a rotating disk discussed earlier
by Ustyugova {\it et al.} (1995, 1999).
       In the reference frame rotating with
the star, the flow velocity  ${\bf v}$
is parallel to
${\bf B}$ at $R=R_{min}$;
that is,
${\bf B \times}({\bf v}-{\bf
\Omega\times  R}) = {\bf0}$ at $R_{min}$.
         We consider the ``free'' boundary
condition for this
velocity, ${\partial ({\bf v} -
{\bf \Omega}\times {\bf R})_R}/{\partial R}=0$.
These two conditions determine the direction
and value of the velocity vector  ${\bf v}$, that is
all three components $(v_R, v_\theta, v_\phi)$.
       The boundary
condition at $R_{min}$ on the magnetic field
has ${\partial(R
B_\phi)}/{\partial R}=0$, while the
       poloidal components
$B_R$ and $B_\theta$ are derived
from the magnetic flux function
$\Psi(R,\theta)$, where $\Psi$
at the boundary is derived from the
frozen-in condition
${{\partial \Psi}/{\partial t}} + {\bf
v}_p\cdot{\bf \nabla}\Psi = 0$.

       At the outer boundary $R=R_{max}$,
we took fixed boundary conditions for all variables
for those cases when  the simulation
region was very large: $R_{max}=55$.
In case of smaller region $R_{max} = 10$
we took free boundary conditions on
all the hydrodynamic variables
and  the $B_\phi$ component
of the magnetic field, if matter
{\it outflows} from the region.
      For the magnetic field we took
${\partial B_\phi}/{\partial R}
= 0$ and ${\partial \Psi}/{\partial t} +
{\bf v}_p \cdot{\bf \nabla}\Psi= 0.$
If matter inflows to the region
($v_R(R_{max})<0$),
then we set this velocity equal to zero.
Results are very similar at these boundary conditions.
    We assume reflection symmetry about the equatorial plane.

\begin{figure*}[t]
\centering
\caption{Distribution of angular velocity in the disk
and corona at different  times $T$
for the same  case as  Figures 4 and 5.
    Only the inner
region of the disk $r < 4$ is shown.
    $\omega$ changes from
   $\omega=0.1$ (white) to
$\omega=1$ (black).
     The solid lines are poloidal field lines
($\Psi=$const lines).
}
\label{Figure 8}
\end{figure*}

\begin{figure*}[b]
\centering
\caption{The figure shows the time
dependence of the mass
accretion rate $\dot M$ and the angular momentum fluxes
carried by matter $\dot L_m$ and
magnetic field $\dot L_f$ to the star
for the case  of a slowly rotating star $\Omega_*=0.19$
which corresponds to $P_*=9.4~{\rm days}$).
The corotation radius is $r_{cor}=3$. }
\label{Figure 9}
\end{figure*}


\subsection {Initial Conditions}

Here we present our new method for
establishing
quasi-equilibrium initial
conditions.
      The star has a magnetic dipole
field with the magnetic moment $\rvecmu$
aligned with
the rotation axis $\hat{z}$,
 with $\Omega_*$  the star's rotation rate.
The magnetic field  is
$
{\bf B}={3(\rvecmu \cdot {\bf R}){\bf R}}/{R^5}
- {\rvecmu}/{R^3},
$
with components in spherical coordinates
$B_R=2 {\mu} \cos{\theta}/R^3$ and
$B_\theta={\mu} \sin {\theta}/R^3$.
       As initial conditions we set up a
low-temperature  $T_d$
disk with a high-temperature $T_c\gg T_d$
corona filling the remainder of the simulation region.
        The disk rotates with
angular velocity close to the
Keplerian value $\omega \approx \Omega_K$.
        The disk extends inward to the radius $R_d$
where the ram pressure of the disk is
equals to the magnetic pressure
of the dipole, $p+\rho {\bf v}^2  = B^2/{8\pi}$.

\begin{figure*}[t]
\centering
\caption{Same as on Figure 9, but for
initial conditions of Type II.}
\label{Figure 10}
\end{figure*}


         In order  to have approximately
equilibrium initial conditions it is
necessary  that the corona initially be
rotating.
       If  the corona initially does not rotate
(e.g., HSM96), or rotates with the angular
velocity of the star (e.g. MS97),
while the disk rotates with Keplerian velocity,
then the coronal magnetic field lines
threading the disk and the star
lead to magnetic braking of the disk
and subsequent
fast accretion of the disk
with a velocity close to free-fall.
       Furthermore, a strong discontinuity
develops in the angular velocity
between the disk and the corona
(except at the corotation radius).
       This leads to the generation of a
strong $B_\phi$ component, while
the dragging of the magnetic field
with the radial inflow of
the disk leads to the generation
of a $B_r$ component.
       This twisting of the field
is  responsible for transient
magneto-centrifugally  driven outflows
(e.g HSM96; MS97; Hirose et al. 1997).
        Such fast evolution may correspond
to periods of violent accretion in
some strongly non-stationary systems.
       However, the star-disk systems
may  exist in more quiescent
configurations which can be studied by
using appropriate initial conditions.

\begin{figure*}[t]
\centering
\caption{Results of simulations of disk-star interaction
for the case of a rapidly rotating star,
$\Omega_*=1$ and $r_{cor}=1$, which
corresponds to $P_*\approx 1.8~{\rm days}$ for the
T Tauri parameters of \S 2.2.
   The density (background) and poloidal
magnetic field lines or $\Psi=$const lines
(solid lines)
   are shown.
      The $\Psi$ contours are exponentially spaced
between $0.15$ and $1$.}
\label{Figure 11}
\end{figure*}

        In order to avoid the strong
discontinuity of $\omega$
and $B_\phi$ on the boundary
between the disk and the
corona, we rotate the initial corona so that
its angular velocity is a constant on a
given cylindrical radius $r=R \sin\theta$
and equal to the  Keplerian rotation
rate of the disk.
       Thus the corona
rotates with different angular
velocities on cylinders with
different radii $r$.
        For such initial conditions we observe much slower
accretion in the disk with no dramatic outflows.
       This  distribution of $\omega$ in the corona
also leads to the twisting of the magnetic field lines.
    Any coronal magnetic field line will
be twisted along its length
by the differential rotation
of nearby layers of corona,
but this twist is distributed along
the length of the field line.
     At the boundary between the
disk and corona the twist is small.
    It is much
smaller than the twist
at the boundary between a rotating disk
and non-rotating corona.
     Differentially rotating corona
also leads to the magnetic braking of the disk,
but it is more gradual and does not lead to a
rapid infall of the matter in the disk.
    Earlier,
a differentially rotating corona
was used as an initial condition in a
study of the
opening of magnetic loops threading
a Keplerian disk, but
   the disk was considered as a boundary
condition (Romanova {\it et al.} 1998).

        Initially, we assume
that $B_\phi=0$ everywhere.
       Also, we assume that
the initial poloidal field is
that of a dipole and that the plasma
is force-free.
       Thus, we search for equilibrium
initial conditions for the disk-corona
system taking into account
gravity, pressure, and rotation.
        To find the initial equilibrium, we
suppose that initially ($t=0$) the
matter is barotropic
$\rho=\rho(p)$.
       In this case one have
(1) ${\bf \nabla} p/\rho = {\bf \nabla} F,$ where
            $F=\int{{dp}/{\rho}};$
(2) the angular velocity $\omega$
depends only on $r=R \sin\theta$;
(3) the centrifugal force has the potential
$\Phi_c=-\int_\infty^r[\omega(r')]^2r'dr'$;
(4) the stationary Euler
equation
  $ ({\bf v \cdot \nabla}){\bf v}+ p/\rho  +
{\bf \nabla} \Phi = 0 $ has the integral:
$$
F +\Phi_c + \Phi = E = {\rm const},
\eqno(5)
$$
  where $\Phi=-GM/R$
is the gravitational potential.

We suppose that the initial
angular velocity of the matter in
the equatorial plane is
$$
     \omega(r, \theta={\pi/2}) = \left\{
\begin{array}{rl}
(k GM/R_d^3)^{\frac{1}{2}}~,  &  r \leq R_d \\
(k GM/r^3)^{\frac{1}{2}}~,  &  r \geq R_d~,
\end{array} \right.
\eqno(6)
$$
where $r=R \sin{\theta}$ is the cylindrical radius,
and $k={\rm const}\sim 1$ is included to take into
account that the disk is slightly non-Keplerian
(in our simulations $k=1.01$).
       Thus, the centrifugal potential in the whole
region is
$$
     \Phi_c(R, \theta)\! = \!\left\{
\begin{array}{rl}
\!\!k (GM/R_d)[1+(R_d^2-r^2)/{2R_d^2}]~, &   r \leq R_d \\
k GM/r~,  &  r \geq R_d ~.
\end{array}
\right.
\eqno(7)
$$

At the boundary
between the disk and corona the
initial pressure is taken to be $p_0$.
       The dependence of the density
on  pressure in the corona and in the
disk can be expressed as
$$
\rho(p) = \left\{
\begin{array}{rl}
\bar{m}p/T_c~,  &  p < p_0 \\ \bar{m}p/T_d~,
&  p > p_0~,
\end{array}
\right .
\eqno(8)
$$
where $\bar{m}$ is the mean particle mass
approximately equal to $m_H/2$.
      We consider that $T_d \ll T_c$.
In the corona, at low values of pressure
and density
$(p < p_0, \rho <\bar{m} p_0/T_c)
$ the plasma  is hot $T=T_c$,
whereas in the disk, at high pressure and
density ($p > p_0, \rho >\bar{m} p_0/T_d$) the
plasma is at a much lower temperature,
$T=T_d$.
       Initially, the boundary between the disk
and corona corresponds to the contact
discontinuity, where $p=p_0$,
$\rho_c=\bar{m}p_0/T_c$  in the corona,
and $\rho_d=\bar{m}p_0/T_d$  in the disk.
In this case
$$
     F = \left\{
\begin{array}{rl}
{(T_c/\bar{m})} {\rm ln}(p/p_0)~,  &
     p\leq p_0 \\ {(T_d/\bar{m})} {\rm ln}(p/p_0)~,
&  p\geq p_0~.
\end{array}
\right.
\eqno(9)
$$

The initial disk-corona boundary in
the equatorial plane
is assumed to be at
$r=R_d$.
    We assume values of $p_0,$ $T_c,$ and $T_d$.
At the point
$(r=R_d,z=0)$,
$p=p_0$, $F(p_0)=0$. Thus, we find $E=-GM/R_d +
kGM/R_d = (k-1)GM/R_d$.
       Then, we can calculate potential $\Phi
+ \Phi_c$ at any point $(R,\theta)$ and find
$F=E-(\Phi+\Phi_c)$.
       Now, we can derive pressure $p$
at any point of the region,
$$
     p  = \left\{
\begin{array}{rl}
p_0 \exp{(F\bar{m}/T_c)}~,  &
     p\leq p_0 \\
p_0 \exp{(F\bar{m}/T_d)}~,
&  p\geq p_0~
\end{array}
\right.
\eqno(10)
$$
and find the distribution of density taking into account
eq. (8).
We considered two  main types of initial conditions which
are described below.

\subsubsection{Initial Conditions of {\it Type I}}

    In the first type of initial conditions,
we place the inner radius
of the disk at $r=R_d=1$.
    We determine the characteristic  density and temperature
in the corona to be $\rho_c=0.01$,  $T_c=1$, and density
in the disk $\rho_d=1$.
    Then we determine the  pressure at the boundary
between the disk and corona, $p_0=\rho_c T_c =0.01$
and then derive
temperature in the disk, $T_d=p_0/\rho_d = 0.01$,
so that the pressure is continuous across the
boundary.
    Then we determine the magnetic field
of the star so as to have the
  magnetic pressure  equal
to the stagnation pressure of the disk matter
at $R=1$.
    This gives
$\mu^2 /{8\pi R_d^6} \approx \rho_d v_K^2$ which
is solved for $\mu$.
    As matter in the disk flows
inward, it forms a funnel flow
at $R\approx 1$.
      We determine $\omega(r)$ from
equation (6) taking
into account that $R_d=1$.
     For $r < R_d$ matter rotates
with an angular velocity corresponding to
$\omega_K (r=1)$, while at $r > R_d$ the
rotation is Keplerian.
    The surfaces of constant $\omega$ are cylinders.
For such a distribution, there are no strong gradients
of $\omega$ in the region.
Then we derive the pressure and density distributions
using equations (5)--(10).

   The left panel of Figure 1  shows the density
distribution in the disk and corona.
    For this  example we took a
relatively small region $R_{max}=7$.
One can see that
the density is high in the disk,
and is $100-200$ times smaller in the corona.
    For these initial conditions the FF starts
close to the initial inner radius of the disk.
    Thus the disk provides a
  ``reservoir" of matter for the FF.
         On the other hand
there is the following disadvantage.
    In most cases, we rotate the
star with an angular velocity $\Omega_*$
  smaller than the initial
angular velocity of matter above the star.
  Fortunately, the
star is observed to  control the rotation
of the nearby regions of  corona in a few time-steps.
We also used initial conditions of Type II,
as described below.

\begin{figure*}[t]
\centering
\caption{Same as on Figure 9, but $\Omega_*=1$ and
$r_{cor}=1$ ($P_*=1.8~{\rm days}$ for the T Tauri parameters
of \S 2.2).}
\label{Figure 12}
\end{figure*}

\begin{figure*}[b]
\centering
\caption{Same as on Figure 9, but $\Omega_*=0.54$ and
$r_{cor}=1.5$ ($P_*\approx 3.3~{\rm days}$
for the T Tauri parameters
of \S 2.2).}
\label{Figure 13}
\end{figure*}

\subsubsection {Initial Conditions of Type II}

In a second type of initial conditions
we place the inner
  radius of the disk $R_d$ at the
corotation radius,
  $r_{cor} = (GM/\Omega_*^2)^{1/3}$.
     Thus for $r < R_d$ both the star and
the corona
rotate with the angular velocity
$\Omega_K(r=R_d)$, while at larger distances the
rotation is Keplerian.
For this initial condition,  $\omega$
varies smoothly
including the variation from the
star and to the corona above it.

An important aspect of the Type II
initial conditions is that the initial
inner radius of the disk is
far from the magnetosphere.
This allows us to  check whether or not
matter moves inward from the corotation
radius.

    The left panel of
Figure 1  shows the initial density
distribution for Type II initial
conditions for $r_{cor}=3$.

\subsection{Hydrodynamic Evolution of the Disk and Corona}

    First, we tested our initial conditions
  with no magnetic
field.
      We observed that for
both types of initial conditions
the disk exists for more
than $300$ rotational periods
$P_0$.
      No high velocities
appear in the disk or corona.
Typical velocities in the
disk are $v_d\sim 0.01-0001$.
    However, our typical time of
simulation of magnetic cases
is shorter: $50 P_0$, so that
we show the result of the evolution
of the disk without a magnetic field
at $t < 50 P_0$.

    We considered three cases, with no viscosity,
$\alpha=0$, and with small viscosities $\alpha=0.01$
and $\alpha=0.02$.
     Figure 2 (a,b) shows that at $\alpha =0$
and boundary conditions of Type I,
the disk's inner edge stays at
$r=1$ during the whole period of evolution.
     The density in the rest of the disk
  slowly redistributes itself in such way that the external
layers move gradually outward.
      The outer regions
of the disk had excess angular momentum  compared to a
stationary $\alpha$-disk (Shakura \& Sunyaev 1973),
and consequently some
matter moved outward carrying away extra angular momentum.
    When we included a small viscosity $\alpha=0.01$
(see panel $c$ of Figure 2), the outer regions
of the disk also moved outward,
but inside the shown region $r < 10$
matter moves inward with velocity $v < 0.0001-0.001$
for $r > 6$, and with larger velocity
$v \sim 0.003-0.006$ for $r < 6$.

     For even higher viscosity
$\alpha=0.02$ (panel $d$ of Figure 2),
the  velocity of the inflow is $v\sim 0.003-0.006$
in the entire disk, $r < 10$.
In both initial conditions we find
that the $\alpha-$viscosity gives a
$\it {slow~ accretion}$ and thus
a good starting point  for our  MHD simulations.
The typical accretion rate is $\dot M\approx 0.05$
for $\alpha=0.01$ and $\dot M\approx 0.1$ for
$\alpha=0.02$.

For the Type II initial conditions  we observed
that for $\alpha=0$
the inner radius of the disk stays
at $r=3$ for $T=50$.
     This shows
that the numerical viscosity is very small
(see panels $e$ and $f$ of Figure 2) compared with
an $\alpha=0.01$ viscosity.
    External regions moved
outward with velocity $v < 0.001$,  redistributing
density and angular momentum.
     For $\alpha=0.01$ and $0.02$, the disk behaved similarly
to one for initial conditions
of Type I:  the equatorial regions
accreted slowly inward with velocity  $v\sim 0.003-0.007$
(panels $g$ and $h$ of Figure 2). The accretion rate
is $\dot M\approx 0.02$ for $\alpha=0.01$ and
$\dot M\approx 0.05$ at $\alpha=0.02$.
    These tests indicate that our
hydrodynamic disk and corona
are appropriate for subsequent MHD simulations of
the disk-star interaction.

\begin{figure*}[t]
\centering
\caption{Disk-star interaction in case of ``torqueless"
accretion at $\Omega_*=0.54$ and
$r_{cor}=1.5$ ($P_*=3.3~{\rm days}$
for the T Tauri parameters
of \S 2.2).
     Contour lines of magnetic flux $\Psi$ are exponentially
spaced between $0.07$ and $0.7$.
       These values of $\Psi$ are smaller
than those in Figure 11 because
density at the inner edge of the disk
is smaller and disk stops at
smaller values of $\Psi$.}
\label{Figure 14}
\end{figure*}

\begin{figure*}[b]
\centering
\caption{Same as on Figure 9, but $\Omega_*=0.67$,
$r_{cor}=1.3$ ($P_0=2.7~{\rm days}$
for the T Tauri parameters
of \S 2.2).
The dot-dashed line corresponds to $\dot L_m$.
     The star spins down most of the time.}
\label{Figure 15}
\end{figure*}
Note, that subsequently we observed that
accretion rate to the star in the magnetic case
 is  larger than in hydro case
(see Figures 9 and 10, where accretion rate
is $\dot M\approx 0.2-0.4$ at initial conditions of
Type I, and $\dot M\approx 0.5-1.3$ at
initial conditions of Type II, which is
connected with small magnetic braking of matter
in the disk, because initial conditions give only
approximate but not exact initial equilibrium.

\section{Disk-Star Interaction for
a Slowly Rotating Star}

        Here, we discuss  simulations of
accretion to a  slowly
rotating star with $\Omega_*=0.19$
($r_{cor}=3$),
which corresponds to a
period $P_*=P_0/\Omega_*\approx 9.4~\rm {days}$ for
a typical  T Tauri star (see \S 2.2).
       The magnetic moment is $\mu=1.06$, which
corresponds to a magnetic field
at the surface of the star
$B_*\approx 1100~{\rm G}$.
        We used initial conditions of
Types I and II to check for the possible dependence
of the evolution on initial conditions.
      The simulations were performed with the grid
$N_R\times N_\theta = 150\times 51$ which has
    $R_{max}=55$.

\subsection{Simulations with Type I Initial Conditions:
Structure of the Disk}

First, we describe results of simulations with Type I initial
conditions.
       We observed that matter
started to move slowly both in
the disk and corona with velocities much
smaller than free-fall.
       No dramatic collapse of the
disk or fast outflows were observed owing to our
quasi-equilibrium initial conditions.

       Figure 3 shows the entire simulation region
    at different  times.
       The outer part of the disk
changes very little during the simulation time $T \leq  50$.
       The magnetic field lines twist and become partially open.
Viewed on this large scale, the disk
is relatively flared, but
only the much smaller region of the disk $R \leq 10$ is
important for disk-star interactions.

Figure 4 shows the evolution
in the smaller region $R \leq 7$.
       One can see, that the inner region
of the disk is influenced by the magnetic field of the star.
For $T \approx 50$ the accretion rate due to
the $\alpha-$ viscosity increased
and magnetic flux was accumulated in the central region.
    We stopped the simulations at
this point, because later the accretion rate increased,
the inner radius of the disk moved closer to the star,
and the case became less interesting for the investigation
of the disk--star interaction.

      Figure 5 shows  the evolution of the
density and poloidal magnetic field
in the region $r \leq 6$.
       The interaction of the disk
with the magnetosphere of the star
led to the reconstruction of the disk inside
the radius which we  call the ``braking"  radius
$r_{br}$.
       In this case $r_{br}\approx 3$.
This radius is an approximate analog of the
radius of the outer transition zone in
the theory of GL79b, which divides areas of
an undisturbed viscous disk and inner regions, influenced by
a magnetic field.
       In our case, the disk for $r > r_{br}$ is
also disturbed by the star's field, but  matter
still moves to this region due to viscosity.
       At smaller radii $r \lesssim r_{br}$, matter is magnetically
braked.
        By coincidence
$r_{br}\approx r_{cor}$.
       The inner regions of the disk
become magnetically braked  after several rotation periods
$P_0$.
          In the region of the magnetic braking
the density in the disk  is $2-4$ times smaller
than in the disk without a magnetic field.
       The inflow speed of the matter
increases to $v\sim 0.01-0.06$.
and matter constantly moves
in the direction of the star.
       The flow towards the
star stops when the matter stress becomes comparable
to the  magnetic pressure of the dipole,
$
B^2/{8\pi} = p + \rho v^2.
$
Note that the main term in the matter stress,
$\rho v_\phi^2$, is due to the disk's rotation.
        We introduce the  plasma parameter,
$\tilde\beta \equiv (p + \rho v^2) / (B^2/{8\pi})$,
which indicates the relative importance
of the matter and magnetic stresses.
     Also, we introduce the magnetospheric radius,
$$
r_m = \frac{\mu^{1/3}}{[8\pi (p+\rho v^2)]^{1/6}}~,
\eqno(11)
$$
which is the radius at $z=0$ where $\tilde\beta=1$.

       Figure 5 shows that matter accumulates
in the region $r\approx r_m$  forming a dense ring.
The density in the ring is
$\sim  2-3$ times larger than the initial density
at the inner radius of the disk.
      At $T = 50$, when more matter comes from the disk,
the density in the ring becomes $\sim 10$ times larger.
       Accumulation of matter near $r_m$ and
formation of a ring in the inner regions of the disk
was also reported by GWB97, GBW99, and MS97, but
these simulations were for  non-stationary conditions.


\begin{figure*}[t]
\centering
\caption{
Disk-star interaction  at different magnetic
momenta of the star $\mu$ after $T=10$ rotations.
Density (background) changes from $\rho=2.5$ (black)
to $\rho=0.004$ (white)
Magnetic flux $\Psi$ (solid lines) is zero at
panel a, and has limits:
$\Psi = 0.1-0.6$ on other panels.}
\label{Figure 16}
\end{figure*}


       For $r \lesssim r_m$ matter moves upward,
away from the equatorial plane
and goes into the funnel flow.
The FF is formed after about one period of
rotation  ($T\sim 1$) and is a stable feature
during  the length of the run, $T=50$.
      Figure 5 shows the $\tilde\beta=1$ line (black line).
The base of the FF  coincides with $r=r_m$; that
is FF starts in the area where the magnetic stress
is equal to the matter stress.
      This is observed in many other cases
with different parameters (see \S 4 \& \S 5).
This is in accord with theories developed in 1970's
(PR72, GL79a,b).
We investigate the physics of FFs in
\S 6.

Matter from the outer
region of the disk $r > r_{br}$
tends to accrete inward.
Often accretion rate is enhanced in the region
where magnetic field has larger radial component
$B_r$ and hence
stronger magnetic braking.

      The accretion rate of matter from the region of
the disk outside the ring
varies with time.
     Consequently, the density in the inner
ring and in the funnel flows also varies with time (Figure 5).
      When more matter accumulates in the ring, then
the base of the FF is closer to the star, and the
magnetospheric radius is smaller.
      For example, at  $T=10$ (panel b of Figure 5),
    $r_m=0.92$; at $T=30$ (panel c)
$r_m=1.2$.
      Subsequently,
for $T > 50$, even more matter came from the disk owing to
viscosity, and the FF moved even closer to the star.
      A stationary state for $T>50$ may exist but
at smaller values of $\alpha$,
$\alpha < 0.02$.

\subsection{Simulations with Type II Initial Conditions}

For Type II initial conditions,
the inner radius of
the disk is taken to be the corotation radius,
$r_{cor}=3$.
      The viscosity parameter is $\alpha=0.02$,
the same as in the above-mentioned simulations.
We were able of observe the evolution for more
than $80$ rotations of the inner radius of the disk.
       Figure 6 shows the evolution
of the inner regions of the disk.
       We observed that the funnel flow did {\it not}
form at the corotation radius
as  predicted in some theories (e.g., OS95).
       Instead, the disk slowly moves inward towards
the star owing to the viscosity (see Figure 6).
      When the disk reached conditions where
    $\tilde{\beta}\approx 1$, the accreting matter
goes up into the funnel flow.
      The FF  formed after $T\approx 10$ rotations.
At longer times,
the structure of the disk and  the FF are
similar to those found for Type I initial conditions
(Figure 5).
       But one
difference is that  for Type II initial conditions
the inward motion of the inner edge of the disk drags
the bases of magnetic loops inward, so that loops of the dipole
are inclined to the disk.
  Such a situation is
possible when the viscosity is larger than the diffusivity
and magnetic flux in the disk accumulates closer
to the star (e.g., S94, OS95), but the rest of the loop
in the corona doesn't move so fast, so that the loops are
inclined to the disk, and the $B_r$ component is significant.
When the magnetic field is inclined away from the
axis of  the disk,
the magnetic force is much stronger than in case with no
inclination (Lovelace, Berk \& Contopoulos 1991,
Oyed \& Pudritz 1997; FE99, FE00).
     We observed opening of the
magnetic field lines close to the
magnetosphere.
     The opening was followed by
reconnection, and subsequent opening. The process was
quasi-periodic with a quasi-period
$T\approx 3.5 ~{\rm days}$.
      We also observed quasi-periodic oscillations of the inner
radius of the disk between radii: $0.7 <r_m < 1.2$.
The nature of
the oscillations is the following: (1) First, the
  loop is stretched  and the magnetic field
``blocks" the path of matter to the FF. (2) Next,
reconnection releases the stress, and  matter accretes
through the FF. (3) The new loop starts to stretch, and the process
repeats. Figure 10 shows the quasi-periodicity of the
accretion rate.
       These oscillations resemble those observed
   by GWB97 and GBW99, but on a much smaller scale.
H97 and MS97 also observed enhanced accretion after
the reconnection of elongated magnetic
field lines near the magnetosphere.
This work supports the idea of
Aly \& Kuijpers (1990) who proposed this mechanism
of quasi-periodic oscillations.

       We tested the dependence of the results on
the viscosity for both types of initial conditions
for  $\alpha=0.01$ and $\alpha=0$. We obtained similar
results, but the FF formed at slightly larger distances $r$
as a result of smaller accretion rate.
Simulations with initial conditions
of Types I and II have shown  examples of
   possible situations which can be realized in nature.
       In the remainder of the paper we use
the Type I initial conditions, because the FF starts earlier
in this case.

\begin{figure*}[t]
\centering
\caption{Close view of a
 quasi-stationary
funnel flow calculated for Type I
initial conditions.
         The color background represents the logarithm of
the density.
     The density has a minimum value
$\rho=0.004$ (blue) and a maximum value $\rho=2$ (red).
         The lines
are poloidal magnetic field lines or equivalently
$\Psi(r,z)=$ const lines.
        The arrows represent the poloidal mass
flux density $\rho {\bf v}$.
       The bold dark line is the field line approximately in
the middle of the funnel flow.
         The variation of different quantities along this field
line are shown in subsequent figures.}
\label{Figure 17}
\end{figure*}


\begin{figure*}[b]
\centering
\caption{Variation of the
density $\rho$, pressure $P$ and temperature $T$ with
distance $s$ from the star's surface
along the fiducial magnetic field line
shown in Figure 17.}
\label{Figure 18}
\end{figure*}

\subsection{Angular Momentum Transport}

      How  is angular momentum transported between the  disk
   and the star?
Does the star spin-up or spin-down?
      In this section we
analyze these questions using the run described in $\S 3.1$.

\subsubsection{Angular Velocity of the Disk}

   We observed that the magnetic field lines which
are responsible for the partial
destruction of the disk at radii $ r < r_{br}$
are also involved in the
transport of angular momentum between the star and the disk.
    The magnetic filed lines connecting
the star and the disk decrease the disk's angular velocity
   in a significant part of the braking region.
Figure 7 shows the
radial distribution of the angular
velocities at different times.
      One can see that the angular velocity
of the disk is significantly
smaller than the Keplerian value in a broad region  $ r < 1.3-1.8$.
     We introduce the radius $r_\Psi$, such that
for $r < r_\Psi$ the magnetic force on the
disk is strong enough
to cause a significant deviation from Keplerian rotation.
    The radius $r_\Psi$  depends strongly on the accretion rate
and varies  between
$r_{\Psi}\approx 1.8$ at  $T=30$,
(when the inner disk is of low density)
and $r_{\Psi}\approx 1$ at $T=10$
(when the inner disk is dense).

Figure 8 shows the spatial distribution
of the angular velocity,
$\omega (R, \theta)$.
One can see that for
$r < r_{br}\approx 3$ the regions of
   constant angular velocity $\omega (R,\theta) = {\rm const.}$
almost coincide with the  regions
of constant magnetic flux $\Psi$.
      Thus, the magnetic field line crossing the disk at radius $r$
rotates in the corona with angular velocity $\omega(r)$.
     This is because most of the corona is matter-dominated,
   $\tilde\beta > 1$ (see black line in Figure 5),
and the disk determines the rotation of the
magnetic field lines.
    Magnetic field lines
which cross the disk in the magnetically-dominated
area (in the FF) rotate with the angular velocity of the star.

\begin{figure*}[b]
\centering
\caption{Variation of
different velocities with
distance $s$ along the fiducial magnetic field
shown in Figure 17.
$v_p$ is the
poloidal velocity (bold line), $v_q$ is the velocity
perpendicular to the
magnetic field line,
$c_{sm}$ and $c_{fm}$ are the slow and fast
magnetosonic velocities, $v_{ff}$
is a local free-fall velocity.
       The dashed line shows the
Mach number ${\cal M}$.}
\label{Figure 19}
\end{figure*}


\begin{figure*}[t]
\centering
\caption{Forces parallel
to the flow
as a function
of distance $s$ from the star's
surface along the fiducial
field line in Figure 17.
      Only part ($\sim 2/3$) of length
of the flow
is shown  in order to
resolve the forces acting near
the disk.
     These
forces  are responsible
for driving matter into the funnel flow.
$F_G$ is the gravitational force,
$F_C$ the centrifugal force,
$F_P$ the pressure gradient
force,  $F_M$ the magnetic force,
and $F_R$ is the resultant force (bold line).}
\label{Figure 20}
\end{figure*}

\subsubsection{Spin Evolution of the Star}

          An important question is the
angular momentum flux to the star.
     To answer this question, we calculated the flux
of angular momentum (about the $z-$axis)
carried by the matter $\dot {L}_m$
and that carried by the `twist' of the magnetic
field $\dot {L}_f$,
$$
\dot {L}_m =
\int {\bf dS}\cdot \rho r v_\phi {\bf v}_p~,\quad
\dot { L}_f=
-\int {\bf d S}\cdot
rB_\phi {\bf B}_p/4\pi~,
$$
evaluated at the surface of the star.
      In addition, we calculated accretion
rate to the  star,
$\dot M =\int {\bf dS}\cdot \rho  {\bf v}$.

      Figure 9 shows the different fluxes
as a function of time.
     Most of the angular momentum flux at
the star is carried by
the magnetic field;
matter carries only $\sim 1\% $
of the total  flux.
     The  flux
$\dot L_f$ is positive so that it
   acts to {\it spin-up}
the star.
     The flux $\dot L_f$
   correlates with
$\dot M$ because incoming matter from
distances $\sim r_m$ is
responsible for bringing in positive angular
momentum.
    At distances $r \sim r_m$ the angular
momentum flux is carried mainly by the
matter.
      These results partially confirm the
hypothesis of
S94, OS95, Li et al. (1996) that matter
should carry  little  angular
momentum as it approaches the star.
       However, our results indicate
that this does not lead to the
``torqueless" accretion predicted by these
authors, because the angular momentum
flux carried by the matter
is  transferred to a flux
carried by the magnetic field
with decreasing distance from the star.
     The small twist of the magnetic field  near the star,
$|B_\phi|/B_p\sim 0.1$ (see \S 6),
is sufficient to carry the
observed angular momentum transport flux
(see discussion  by Wang 1997).

      From our simulations we
find that the magnetic field lines
responsible for spin-up or spin-down
of the star pass through the disk at
distances $r < r_{\Psi}$.
     Consequently the spin evolution of
the star depends on
the location of the corotation radius $r_{cor}$
relative to $r_\Psi$.
    A star may spin-up, spin-down,
or be in the regime of the ``torqueless" accretion
(see \S 4.2).
For the case considered in this subsection of a slowly
rotating star $r_{cor}=3$,
the star spins-up.

\begin{figure*}[t]
\centering
\caption{Angular velocity $\omega$
and the ``twist'' $|B_\phi|/B_p$ as a function
of distance $s$ from the star's
surface  along the bold
field line in Figure 17. }
\label{Figure 21}
\end{figure*}

\section{Accretion to Rapidly Rotating Stars}

     We performed a set of simulations
for  different angular velocities of the
star $\Omega_*$.
     Simulations were done for Type I initial
conditions  and with viscosity $\alpha=0.01$.
      The grid was $N_R\times N_\theta=100\times 51$
corresponding to the smaller region $R_{max}=10$.
     The fastest  rotation considered
was such that
at $r_{cor}=1$, $\Omega_*=1$ which corresponds
to a rotation period $P_0=1.8~{\rm days}$
for the T Tauri  parameters of \S 2.2.
(The  conversion formulae for dimensional period is
   $P_*=1.8/\Omega_*~{\rm days}$).
    In this section we discuss the dependence
   of the spin-up/spin-down rate of
the star  on $\Omega_*$.

\subsection{Rapidly Rotating Star: Spin-Down}

      Here we discuss the case of a
rapidly rotating star,
$\Omega_*=1$, $r_{cor}=1$ ($P_*= 1.8~{\rm days}$),
where $r_{cor}\approx r_m$.
     We observed that initially the accretion disk
moved outward because the magnetic field
of the star transferred its
angular momentum to the disk.
    Later, for $T >3$ the disk started
to move inward towards the star,
and finally a funnel flow formed
(see Figure 11).
     In this case the thickness of the
inner part of the disk is larger
than that in the case of slowly rotating star, and matter
accretes preferentially along
the top layers of the disk.

Figure 12 shows  the observed fluxes of matter
and angular momenta to the star.
      Note, that the flux of an angular momentum
is carried mainly by the  magnetic field,
$\dot L_f$, which is negative so that the
star spins down.
      Thus, we observe that as matter
{\it accretes} to the star angular
momentum flows out from the  star.
     As in the case of slowly rotating star, we
observe a correlation between the matter
and angular momentum fluxes.

\subsection{What is the Value of $\Omega_*$ for
Torqueless Accretion?}

We arranged a set of simulations
to test if a particular value of
$\Omega_*$ gives ``torqueless'' accretion.
        According to GL79b, at some angular
velocity of the star
$\Omega_{crit}$, positive
magnetic torque associated with the region
$r < r_{cor}$ cancels the negative magnetic
torque associated with
the region $r > r_{cor}$
and the star can accrete without changing
its angular momentum.

We performed simulations
at different $\Omega_*$
(corresponding to $r_{cor}=1.2, ~1.3, ~1.4,
1.5, ~1.7, ~2,~ 3,$ and  $10$)
and observed, that for
$\Omega_*\approx 0.54$ ($r_{cor}=1.5$)
the angular momentum flux $\dot{L}$ wanders back and
forth around zero (see Figure 13).
       That is, it is positive for
some time ($\sim 10P_0$),
then becomes negative for a similar length of
time, and
then become positive again.
    We conclude that this $\Omega_*$
  corresponds to torqueless accretion.
     The explanation for torqueless accretion
was proposed in the  $1970$s (e.g., GL79b).
     Namely, when more matter is supplied to the disk,
the magnetospheric radius $r_m$
moves closer to the star, and as a result
there is a positive angular momentum flux to the star.
     When less matter accretes, the magnetospheric
radius $r_m$ moves outward, and the
flux of angular momentum is negative.
      Figure 14 shows the evolution of the
density and poloidal magnetic field   for
times corresponding to
positive angular momentum flux to
the star (left panels), zero
angular momentum flux (middle panels)
and negative angular momentum flux (right panels).
     The  variations of the accretion rate may be
connected with accumulation of matter at
$r \sim r_{br}$ where  in this  case
$r_{br}\approx 3 - 3.5$ and subsequent accretion through
instabilities.

       The ratio of the corotation radius $r_{cor}$
   to the magnetospheric radius $r_m$ is of general interest.
At the moments when the torque is zero,
$\dot L=0$ (see Figure 13), the
magnetospheric radius had the
following values:
at $T=24$, $r_m=1$; at $T=35$, $r_m=1.07$; at $T=46$, $r_m=1.1$.
That is, for $\dot{L}=0$ the ratio is:
$r_{cor}/r_m\approx 1.4 - 1.5$.
     For similar simulations for faster
rotation of the star
$\Omega_*=0.67$ ($r_{cor}=1.3$),
the angular momentum flux  also wandered around
zero value (see Figure 15), but most of time it
was negative so that the star spins down.
     For simulations with more slowly rotating
stars or larger corotation radii,
$r_{cor} > 1.5$, ($r_{cor}=1.7,~ 2,~ 3,~ \&~ 10$),
we observed that the star spins up.

\section{Dependence on the Stellar Magnetic Field}

We did simulation runs for  different
magnetic moments,
$\mu=0,~ 0.21,~ 0.42,~ 0.63$ and $1.1$, with
other parameters fixed, $\Omega_*=0.19$ ($r_{cor}=3$), and
$\alpha=0.01$.
     At $T=0$ the inner radius at of the disk
was at $r=1$ in all cases.
      The grid was similar
to the one used in $\S 4$.

Figure 16 shows result after $T=10$ rotations.
One can see that at smaller values of  $\mu$,
the inner radius of the disk and FF are settled
closer to the star than in the case of larger
$\mu$. In all cases, we observed that the
base of the FF approximately
coincides with the magnetosphere radius $r_m$, but at smaller $\mu$
this radius is smaller in accord with equation (11).

  In case of even smaller $\mu$ the direct accretion
to the surface of the star is observed, like in simulations by
  HSM96 and in some simulations by MS97.

The cases with stronger magnetic moment $\mu = 1.1, 0.63$
are more appropriate
for explanation of T Tauri stars, because the gap between
the star and the disk is comparable or larger
than the radius of the star. At smaller magnetic momenta
$\mu =0.42, 0.21$, the gap is too small.

\section{Physics of  Funnel Flows}

     Here, we discuss  the physics
of a typical funnel flow,
the case of a slowly rotating star with
$r_{cor}=3$ (\S 3) after  $T=30$ rotations.
      The specific heat ratio is $\gamma=5/3$.
    Figure 17 shows an enlarged picture
of the FF.
         We determined the variation of different
parameters as a function of distance ($s$)
from the surface of the star
along a fiducial poloidal field line through
the middle of the FF.
      This field line is shown by the
thick black line in Figure 17.
      Figures 18 -- 21 show the variation of different
parameters along this field line.
    Figure 18 shows that the density $\rho$ and pressure $P$
   decrease along the FF, then increase again near
the surface of the star.
The temperature $T$ initially varies slowly, but
increases strongly approaching the star as a result of gas
compression.

         Figure 19 shows the
variation of different velocities
along the FF.
    Matter flows along the FF with
poloidal velocity $v_p$.
     The velocity perpendicular
to the poloidal magnetic field,
$v_q$, is very small, $|v_q| << v_p$.
        The poloidal velocity $v_p$ increases and
becomes larger than the slow magnetosonic velocity
$c_{sm}$ at $\sim 0.4$ of the way.
      This point is expected
to be closer to the disk in the case of
cooler (and thinner)
accretion disks (Koldoba et al. 2002).
     The flow become
supersonic with the Mach number reaching ${\cal M}\approx 3.6$
at the surface of the star. Near the star, the poloidal
velocity is close to the free-fall velocity:
$v_p\approx 0.7 v_{ff}$.

         The poloidal velocity of the FF
is much smaller than the Alfv\'en velocity $v_A$.
The ratio decreases from
$v_p/v_A\approx 1/3$ in the middle of
the FF, to $v_p/v_A\approx 0.07$ at the star.
Thus, the flow is strongly sub-Alfv\'enic.
        This validates the analysis of
Koldoba {\it et al.} (2002)
which proposed that the funnel flows
were sub-Alfv\'enic.

       It is important to understand the
   forces which drive matter out of
the equatorial plane of the disk into
the funnel flow, and which
forces accelerate the flow
towards the star.
       The force per unit
mass $F$ along the magnetic field line
is obtained by multiplying
the Euler equation by
${\bf B}_p/|{\bf B}_p|$.
       This gives
$$
F = \omega^2 r \sin \chi -
\frac{1}{\rho} \frac{ \partial p}{
\partial s} - \frac{\partial \Phi}{\partial s}
-{1\over{8 \pi \rho
r^2}} {\partial(rB_\phi)^2\over\partial s}~,
\eqno(12)
$$
(Ustyugova {\it et al.} 1999),
where $\chi$ is the
angle of the field line to the $z-$axis.
       The terms on the right-hand-side
represent  the centrifugal
force, the pressure gradient force, the
gravitational force, and the magnetic force.

Figure 20 shows
forces in the outer two-thirds
of the poloidal path of the FF.
     We see that the pressure gradient force
is responsible for driving matter up and
out of the disk and into
the funnel flow, while the gravitational force is
responsible for the acceleration of
matter in the rest of the FF.
          Note that the
magnetic force in equation (12)
is small and  acts in the opposite direction.
        Thus, we did not observe
the ``magnetic levitation''
force predicted by Li \& Wilson (1999).
       We observed a finite $B_\phi$ field
above the disk in the FF
(see Figure 21), but the magnetic force
is much smaller than the other forces and
it is not sufficient to levitate
matter above the disk.

     Figure 21 shows that the
angular velocity of the
funnel flow decreases gradually
from its value at
the inner edge of the disk
($r=R_d$) to a small negative value
at the star.
     The negative value of $\omega$ at the star
is due to the fact that the magnetic field
has a small inclination in the direction of
rotation of the star.
     For this reason matter moving
along the magnetic field line comes to the star
with an angular velocity less than $\Omega_*$
(see also Ghosh et al. 1977; Muzerolle et al. 2001).

        Figure 21 shows the variation
of the toroidal magnetic field along
the funnel flow.
        The twisting of the dipolar field
is relatively small with the ``twist''
   $\gamma_\phi \equiv |B_\phi|/B_p < 0.15$.
At the star, $\gamma_\phi \approx 0.04$.
     This is in accord with estimations by
Wang (1997), who has shown that
the twist near the star should be small, but not
a zero.
       Away from the funnel
flow $|B_\phi|$ is
of the order of $B_p$.
        We never observed conditions where
$|B_\phi| >> B_z$ (discussed, e.g., by
GL79a and  Agapitou \& Papaloizou 2000).
The reason is that  for $|B_\phi| >> B_p$, a torsional
      Alfv\'en wave forms and propagates
outward into the corona where it dissipates.

In reality,  matter may
accrete through the magnetosphere
owing to $3D$ instabilities, for example,
the Rayleigh-Taylor
or Kelvin-Helmholtz instability (Arons \& Lea 1976a,b,
Spruit \& Taam 1990).
     Realistic investigations of the role of
these instabilities is possible
with $3D$ MHD simulations.

\begin{figure*}[t]
\centering
\caption{Density and poloidal magnetic
field lines for a case with outflows.
     The magnetic moment of the star $\mu=5.3$
is $5$ times larger than that
   used in most of our simulations.
      The density in the corona is $3$ times
smaller than in other simulations.
Color background represents the density, white
solid lines - contours of  magnetic flux, black solid line
is the line $\tilde\beta=1$.
      Level lines
of density are exponentially distributed
between the minimum,  $\rho=4\times 10^{-4}$, and
the maximum,
$\rho=17.6$ values. The $\Psi$ lines are
exponentially distributed between $\Psi=0.1$
and $\Psi=3$. Arrows are  velocities. The grid used
was
$N_R\times N_\theta=110\times 71$, and
the viscosity parameter was $\alpha=0.01$.}
\label{Figure 22}
\end{figure*}

\section{Opening of Coronal Magnetic Field
and Outflows}

     In our simulations
we observed  coronal
   magnetic field lines at
large distances from the star opening
due to the initial differential
rotation.
    Closer to the star the  field
lines were mostly closed,
contrary to the predictions by
Aly (1986), LRBK95 (see also
Newman, Newman, \& Lovelace 1992;
Livio \& Pringle 1992;  Lynden-Bell \& Boily 1994;
Miki\'c \& Linker 1994, Amari et al. 1996;
FE99; FE00; Ustyugova et al 2000;
Uzdenskyi, K\"onigl \& Litwin 2002).
      The difference in behavior can be
explained as follows.
     The analytical or semi-analytical
models  suppose that the inertia
of the of the  matter is very small,
$\beta=8\pi p/B^2 << 1$,
so that the plasma is
in the ``coronal" limit of Gold and Hoyle (1960).
	    However,  the inertia
of the matter may be important.
     Romanova et al. (1998) simulated
the dynamics of magnetic
loops in the corona of a differentially
rotating disk where
$\beta\sim 1$.
    They observed that
in the magnetically dominated regions
($\beta < 1$) the magnetic
loops opened whereas
in the matter-dominated regions ($\beta > 1$)
the loops did not open.

       The energy-density of the dipole field
decreases rapidly with
distance ($\sim R^{-6}$) compared
with the matter
energy-density in our simulations which varies
slowly within  the computational
region.
    For this reason the magnetically-dominated region is
restricted to a relatively
small volume, $R < 1-1.5$, around the
star (see black line at Figure 5).
    To overcome this feature of
the dipole field, GWB97 and GBW99
adopted a rapid fall-off of density in the corona,
$\rho\sim R^{-4}$.
     For this dependence the region of
small $\beta$ is greatly expanded and
opening of the magnetic field is favored.

     In order to investigate
possible stronger outflows, we chose
   parameters so as to increase the region
of magnetically-dominated corona.
       The magnetic moment was increased $5$ times,
$\mu=5.3$ (which corresponds to a
magnetic field  $B=5.3 ~{\rm kG}$ at the surface of the star).
      The density in the corona was decreased
$3$ times, $\rho_c=0.003$.
    This led to a
   higher Alfv\'en speed near the star, so that we
increased the inner radius
$R_{min}=0.8$ to cover the region of high Alfv\'en
velocity, and increased the resolution of the grid to
$N_R\times N_\theta=120\times 71$.
     We have $\tilde\beta=1$
at $r=R_d=1$ as in the other runs
described in this work.
      The magnetic field is stronger and
the  initial density in the
disk is about $10$ times higher
than in other simulations.

     Figure 22 shows the results of the simulations.
We observed that magnetic braking led to accretion of
the inner part of the disk
with velocity $v_r\approx 0.01-0.1$.
     Thus matter  in the disk concentrated in
a dense ring with $\rho\sim 10-14$.
    The disk angular velocity is
significantly smaller than the Keplerian
value but larger than the
angular velocity of the star, $\Omega_*=0.19$,
which is a result of the interaction with the
  strong stellar
magnetic field.
     A significant region above the corona corotates
with the disk.
   Analysis of forces, similar to that of \S 6
shows that in the region $r < 2$, the centrifugal force
drives matter along the magnetic field lines, but the
force is not strong enough to open the magnetic field lines
and matter accretes to the star forming multiple funnel flows
(see Figure 22).
     However,
the outer magnetic field lines, which are
located at the edge of the magnetically-dominated region,
become gradually  open owing to
combination of centrifugal and magnetic forces.
     Matter accelerates along the region of open field lines
up to velocities $v\sim 0.5 v_{esc}$,
where $v_{esc}$ is the escape velocity.
      We anticipate that
if the density in the corona
decreases faster than in our model,
then matter will outflow from the region.
     The observed  matter and angular momentum fluxes
of outflowing matter are small.

    This test simulation provides evidence
   of how outflows can be
formed from the disk in the quasi-stationary regime.
Subsequent research is needed in this direction at different
distributions of density of matter in the corona.

\section{Application to T Tauri Stars}

         Our simulations are well suited for
modelling  funnel flows to  T Tauri
stars  because the entire flow can be well
resolved.
    In the observed systems, the ratio of the
inner radius of the disk
where the FF begins to the
stellar radius is $R_d/R_* \sim 2-4$.
Such conditions are realized in our simulations.

     The detailed nature of the
magnetic fields of T Tauri stars
is not known, and the fields may differ
significantly from a dipole field (e.g., Safier 1998).
      However, recent spectral studies
of T Tauri stars  show
   broad
emission lines which
   give  evidence of matter
inflow  with nearly free-fall velocities
(e.g., Hartmann et al. 1994; Lamzin, et al. 1996;
Muzerolle et al. 2001).
      Such velocities may be easily explained
in the model where matter inflows supersonically
in a funnel flow along a dipolar field.

     In the following we perform numerical estimates for
the funnel flow
to a typical T Tauri star using
conversion scheme of \S 2.1.
     We assume the inner radius
of the disk is $R_0 = 2.8 R_*$,
the star's mass
$M_{0.8}\equiv M/(0.8 M_\odot)$, its
radius $R_2\equiv R/(2 R_\odot)$, and its surface
magnetic field $B_3=B/(10^3 ~{\rm G})$.
      Using these values
we can derive reference values for
the different physical variables.
     For the flow velocity,
$v_0=(GM/R_0)^{1/2} \approx
1.63\times 10^7 M_{0.8}^{1/2} R_2^{-1/2}~{\rm cm/s}$.
      For the period of rotation at $r=R_0$,
$P_0=2\pi R_0/v_0 \approx
1.78~M_{0.8}^{-1/2} R_2^{3/2}~{\rm days}$.
       For the density, $\rho_0=B_0^2/v_0^2\approx
6.93\times 10^{-12} B_3^2 R_2 M_{0.8}^{-1}~{\rm g/cm^{-3}}$.
      For the accretion rate, $\dot M_0=\rho_0 v_0 R_0^2 \approx
2.8\times 10^{-7} B_3^2 M_{0.8}^{-1/2} R_2^{5/2} ~{\rm M_\odot/year}$.
     For the temperature of the corona,  $T_0\approx 1.07\times 10^6
M_{0.8} R_2^{-1}~{\rm K}$.

      In the following  we calculate the
different physical  variables
along the middle of the funnel
flow shown in Figure 17.
       From Figure 19 the
matter reaches a maximum poloidal
velocity $v_{p*}\approx 277~{\rm km/s}$
at the star's surface, which is $79\%$
of  the free-fall  velocity
$v_{ff}\approx 350~{\rm km/s}$.
      The funnel flow will of course
give a distribution of velocities along
the line of sight to a distant observer.
     The smallest line of sight velocities
are for matter immediately above the disk
where the speeds are  $ \sim 10\%$ of
$v_{p*}$.
     The line of sight velocity distribution
is important for the interpretation of
observed spectral lines
( e.g., Folha and Emerson 2001).
    The poloidal velocity is $\sim  1/3$
the Alfv\'en velocity $v_A$ above the disk (at $s\approx 1$)
and decreases to $v_p\approx  v_A/15$ at the
star's surface.
      The velocity across the magnetic field
lines $v_q$ is very small (see Figure 19).
     Thus, matter flows along the magnetic field.
The flow become supersonic above the disk and the Mach
number reaches ${\cal M}  \approx 3.6$ at the surface
of the disk.

       The initial density of
matter in the disk at $R_0$ is
$\rho_0\approx 6.9\times 10^{-12}~ {\rm g cm^{-3}}$,
and the density of matter in the FF (see Figure 18)
varies in the range
$(1.2 - 9.2)\times 10^{-12}~
{\rm cm^{-3}}$ which is
typical for T Tauri stars.
        Our typical accretion rate to the star
through the FF  is
$\dot M \sim (0.6 - 1.2)\times 10^{-7}~
{\rm M_\odot}/{\rm year}$, which agrees
with estimates for T Tauri stars
(Hartmann {\it et al.} 1998).
      The initial temperature in the disk is
$T\approx 9.6\times 10^3~{\rm K}$
which is several times
higher than expected in the
disks of T Tauri stars at $R_0$.
        As mentioned earlier this higher
temperature is needed to give adequate
resolution in the vertical direction
through the disk.
             We observed that matter in
the FF was heated by adiabatic
compression to  higher temperatures,
$T\sim  1.2\times 10^5~{\rm K}$.

        The fact that we use
   ``free'' boundary
conditions at the surface of the star
means that we cannot account for the
stand-off shock at the stellar surface,
surface heating, and other physical
phenomena close to the star.
       Study of these processes can be
done with a separate set of simulations.

The investigated  quasi-stationary funnel flows
correspond to Classical T Tauri stars
(CTTS) (see Hartman 1998) where variability
is not strong and possible outflows are weak.
    Previous simulations of very fast
accretion with subsequent strong outflows
were done by
HSM96, MS97, H97, GWB97, and GBW99, and these
may correspond to earlier stages of evolution
of young stellar objects
(YSOs of classes 0 and I)
which show evidence of episodic non-stationary
accretion
(e.g., FU Orionis type stars) and strong outflows
(e.g. HH30).
      Non-stationary accretion may be initiated
by thermal instability (Lightman \& Eardley 1974)
or by global magnetic instability
of the disk (Lovelace,
Romanova, \& Newman 1994; Lovelace, Newman, \&
Romanova 1997), or by tidal interaction with a binary
companion (Larson 2002).

\section{Conclusions}

\subsection{Funnel Flows}

      We have done a wide range of
MHD simulations of disk accretion to
a rotating aligned dipole in order to
understand the different accretion phenomena.
    The simulations show that funnel flows (FF), where
matter flows out of the disk plane and
essentially free-falls
along the stellar magnetic field lines,
are a robust feature of disk accretion to
a dipole.  Specifically, we find that

~(1).~ The disk truncates
and a funnel flow forms
near the magnetosphere radius $r_m$,
  where magnetic
pressure of the dipole is
comparable to the kinetic plus thermal
pressure of the disk matter.

~(2).~ The velocity of matter in the FF
is much smaller than the Alfv\'en velocity,
$|{\bf v}| \sim (0.05-0.3 ) v_A$, so that
matter flows along the magnetic field lines.
  The funnel flow accelerates and become
supersonic. The Mach number is ${\cal M}\approx 3-4$
at the surface of the star.
At the star velocity  is close to the free-fall velocity:
$v_p\approx 0.7 v_{ff}$.

~(3).~ The angular velocity of the FF
gradually varies from
its value at the inner edge
of the disk
   to the angular
velocity of the star.
        The `twist' of the magnetic
field lines in the
FF is small, $|B_\phi|/B_p < 0.1$,
and it has a maximum
approximately in the middle of the FF.

~(4).~The main forces which are responsible for
dragging matter to the FF are matter pressure gradient force
(near the disk) and gravitational force in the rest of the FF.
Magnetic force is negligibly small.

~(5).~About $1/3$ of the magnetic flux
responsible for the spin-up/spin-down the star
goes through the FF, while the
remainder is above the FF.

\subsection{Disk-Star Interactions}

     Regarding the interaction between
the disk and the star we find that

~(1)~ The magnetic field of the star influences the
nearby regions of the disk inside a radius
$r_{br}$, while viscosity dominates at
larger radii.
     The radius $r_{br}$ depends on
magnetic moment of the star $\mu$
and density in the disk.

~(2).~ Inside the radius $r_{br}$  the disk
is strongly inhomogeneous. The density is $2-3$
times smaller than in the disk without magnetic field.
Magnetically braked
      matter accumulates near magnetosphere  and
forms a dense ring and funnel flow.

~(3).~   The star may spin-up or spin-down depending
on the ratio of its rotation rate to the rotation
rate at the inner radius of the disk.
     We find that ``torqueless" accretion is
possible when $r_{cor}/r_m\approx 1.5$, where
$r_{cor}$ is the corotation radius.

~(4).~ At the star's surface,
the angular momentum flux is transported  mainly
by the `twist' of the magnetic field.
     Angular momentum carried by matter in the disk
at $\sim r_m$
is transferred almost completely to the
magnetic field at the star's surface.

~(5).~ The coronal magnetic field
is observed to open and
close, but strong outflows  were not
observed for the considered
parameters and quasi-equilibrium initial conditions.
     In the area of the disk where the
field is strong, $r < r_{\Psi}\approx  0.5 r_{br}$, the magnetic field
lines tend to decelerate/accelerate rotation of the disk
instead of being opened.
Besides, opening of magnetic field loops is suppressed by matter dominated
corona compared to GBW99 who accepted dependence $\rho\sim R^{-4}$.

~(6).~ Strong outflows may be associated
with strongly non-stationary
accretion in the disk
as  observed in simulations
by HSM96, MS97, and H97
   or when the disk is very dense
and magnetic field lines are inclined to the disk
as in simulations by FE99, FE00.
      Alternatively the outflows
may occur in cases where the density of the
corona decreases sufficiently rapidly with distance
as in the simulations by  GWB97 and GBW99.
        Sporadic outflows can arise from a
global magnetic instability of the disk
(e.g., Lovelace {\it et al.} 1994, 1997, 2002),
but they may be absent during the quiescent
evolution of the disk-star system.

\acknowledgments
This work was supported in part by NASA grant
NAG5-9047, by NSF grant AST-9986936.
      RMM thanks NSF POWRE grant for partial support.
RVEL was partially supported by grant NAG5-9735, GVU
and AVK were partially supported by INTAS grant
and by Russian program ``Astronomy.''
      The authors thank the referees for
numerous comments and suggestions.



\end{document}